# Effects of Nanofeatures Induced by Severe Shot Peening (SSP) on Mechanical, Corrosion and Cytocompatibility Properties of Magnesium Alloy AZ31

Sara Bagherifard[a], Daniel J. Hickey[b], Stanislava Fintová[c], Filip Pastorek[d], Ines Fernandez-Pariente[e], Michele Bandini[f], Thomas J. Webster[b], and Mario Guagliano[a]

[a]Department of Mechanical Engineering, Politecnico di Milano, Milan, Italy
[b]Department of Chemical Engineering, Northeastern University, Boston, MA, USA
[c]Institute of Physics of Materials, Czech Academy of Sciences, Brno, Czech Republic
[d]Department of Material Engineering, Faculty of Mechanical Engineering, University of Zilina, Slovak Republic
[e]Department of Materials and Metallurgy Engineering, University of Oviedo, Gijón, Spain
[f]Peen Service S.r.l., Bologna, Italy

**Abstract**
The application of biodegradable magnesium-based materials in the biomedical field is highly restricted by their low fatigue strength and high corrosion rate in biological environments. Considering the sensitivity of both fatigue strength and corrosion rate to the surface characteristics of the material, apposite surface treatments could address these challenges. As a low cost and versatile severe plastic deformation technique aimed at inducing surface grain refinement, severe shot peening has been effective in enhancing mechanical properties and promoting cellular interactions on non-degradable biocompatible metallic materials. Herein, we treated the surface of a biocompatible magnesium alloy AZ31 by severe shot peening in order to evaluate the potential of surface grain refinement to enhance its functionality in a biological environment. The AZ31 samples were studied in terms of a wide variety of micro/nanostructural, mechanical, and chemical characteristics in addition to cytocompatibility properties. The evolution of surface grain structure and surface morphology were investigated using optical as well as scanning and transmission electron microscopy. Surface roughness, wettability and chemical composition, as well as in depth-microhardness and residual stress distribution, and corrosion resistance were investigated. Successive light surface grinding was used after severe shot peening to eliminate the effect of surface roughness and separately investigate the influence of grain refinement alone. Cytocompatibility tests with osteoblasts (or bone forming cells) were performed using sample extracts. Results revealed for the first time that severe shot peening can significantly enhance mechanical properties without causing adverse effects to the growth of surrounding osteoblasts. The corrosion behavior, on the other hand, was not improved by severe shot peening; nevertheless, slight grinding of the rough surface layer with a high density of crystallographic lattice defects, without removing the entire nanocrystallized layer, provided a good potential for improving corrosion characteristics after severe shot peening and thus, this method should be studied for a wide range of orthopedic applications in which biodegradable magnesium is used.

## 1. Introduction

A major challenge in the use of most metallic materials, such as titanium alloys and stainless steels, which are commonly used for load-bearing skeletal implants, is their non biodegradability when used as fixation plates, pins, rods, screws, etc. Once these stabilizing implants accomplish their mission, they become permanent foreign objects that can be prone to late-stage complications, such as wear debris, bone loosening, infection, as well as overall a permanent site for inflammation. The aforementioned materials also exhibit a notable mismatch of mechanical characteristics with natural bone, which can initiate stress shielding. Stress shielding results in decreased bone density due to a lack of stimulus for the continual remodelling necessary for maintaining bone mass [1] and can cause further complications, such as a delayed union, implant migration, subsidence and pseudo-arthrosis [2, 3]. Hence, identifying an implant material with mechanical characteristics comparable to those of mammal bone tissue that can stabilize a damaged bone segment under relatively large amplitudes of static and dynamic loading over time, and then resorb without adverse tissue reactions, can be an attractive strategy to address all of the aforementioned issues.

Magnesium (Mg) based materials with bio-resorbable non-toxic degradation products exhibit a high strength to weight ratio, and mechanical characteristics closer to bone, when compared to other metallic materials currently in clinical use. Mg has a density of 1.74 gcm$^{-1}$ (bone: 1-2 gcm$^{-1}$, Ti alloys: 4.5 gcm$^{-1}$, 316L: 8 gcm$^{-1}$) [4], an elastic modulus of 41-45 GPa (bone: 3-20 GPa, Ti alloys: 110-117 GPa and 316L: 193 GPa) and an ultimate tensile strength of 160-263 MPa (bone: 1.5-283 MPa, Ti alloys: 550-860 MPa and 316L: 490 MPa) [5, 6]. Besides, Mg has been reported to be among the most biocompatible metals, indispensable for many biochemical processes in the human body, aiding over 300 enzymatic reactions that help maintain the functions of muscles, nerves, regulate heartbeat, blood sugar level, and the immune system [7]. There are also reports on the positive effect of Mg ion concentration on enhancing osteoblast adhesion and proliferation [8-11] and possible improved osteoblast functions on Mg with nanoscale surface features [12, 13]. However, fast and uncontrolled degradation of Mg alloys and excessive hydrogen gas production in the physiological environment has fundamentally limited their application in the biomedical field, despite their long list of favorable characteristics [14]. Unpredicted corrosion can result in a rapid loss of structural integrity and premature failure before sufficient healing of the adjacent tissue is achieved.

Another drawback of Mg alloys is their inadequate fatigue properties especially under the high cycle fatigue regime [15-17]. There are some reports of using Mg alloys for cardiovascular stents [18] and some recent studies on developing Mg based orthopaedic implants [19], but their application for load bearing implants still remains a challenge. Various solutions have been put forward to enhance their corrosion resistance, including alloying [20], surface coatings [21, 22], and a wide variety of other surface treatments such as thermal, hydrothermal and alkaline heat treatments, chemical and electrochemical surface modifications [23], etc. However, many of these approaches can cause other complications including a risk of compromising biocompatibility and mechanical characteristics or delamination and inadequate bonding of the coating upon which its protecting function can be impaired. Thus, the quest to find efficient methods for enhancing the mechanical and biodegradation properties of Mg alloys is still challenging and on-going.

More recently, researchers have investigated the effect of grain refinement of Mg-based materials through the application of severe plastic deformation (SPD) techniques. SPD techniques result in extreme grain refinement through large plastic deformations applied at high strain rates and relatively low temperatures causing evident enhancement in several mechanical properties and global performance [24-27]. In this regard, there have only been a few studies on the application of SPD based approaches on different Mg alloys, which report enhanced mechanical properties [28-31]. Nevertheless, in terms of corrosion and degradation resistance of Mg alloys after SPD, there are many contradicting observations. Lower corrosion current density and a nobler corrosion potential were reported for ZE41A after equal channel angular pressing (ECAP) [32]. ECAP-processed AZ31 also confirmed that the highest corrosion resistance was obtained on samples with the smallest grain size after long periods of immersion in phosphate-buffer solution (PBS) [33].

The results of another study on the degradation characteristics of hot rolled and ECAP processed AZ31 in Hank's solution, showed later degradation for hot rolled samples (grain size of 15 µm) compared to the base material (grain size of 450 µm), while ECAP (grain size of 2.5 µm) did not affect degradation rate [34]. Ultra-fine grained AZ31 obtained by ECAP exhibited improved corrosion characteristics mainly after long exposure times. Polarization resistance after 5 minutes in 0.1 M NaCl was equal to the not treated and ECAPed alloy but it was doubled for the ECAPed samples after 168 hours compared to the base material [35]. ECAP was reported to reduce localized corrosion through decreasing the size of the second phase particles and promoting microstructural homogeneity on ZK60 and pure Mg but showed a minor effect on the general corrosion resistance [36].

Surface mechanical attrition treatment (SMAT) applied to pure Mg and Mg-1Ca alloy, was reported to result in enhanced hardness and wettability but reduced corrosion resistance [37]. However, treatment of AZ91D magnesium alloys by SMAT resulted in contrasting results on corrosion behavior depending on processing parameters. SMAT treatment using 2 mm balls resulted in a 50 μm affected surface layer depth with an average grain size of 11 nm and significant improvement of corrosion properties. However, using 5 mm balls led to a significant decrease in corrosion properties, even though smaller grain size (8 nm) and greater thickness of the affected layer (150 μm) was obtained. Increased defect density during the latter treatment was determined as the cause of the deteriorated corrosion properties [38]. Some researchers report dependency of electrochemical test results on the applied plastic strain and the induced grain size for AZ31 [39]. Another study emphasized the greater effect of the crystallographic orientation and basal texturing on defining corrosion characteristics of AZ31B treated by severe plasticity burnishing (SPB) when compared to the effect of grain refinement. They reported a remarkable influence of crystallographic orientation on the corrosion resistance, whereas grain sizes smaller than 1 μm were reported not to affect the corrosion characteristics [40]. With the broad scatter of research in terms of material, grain refinement techniques, grain size range, and test conditions, systematic studies are still required to address the fundamental gaps in determining the effects of grain refinement on the mechanical and corrosion properties of Mg-based alloys.

In the present study, we have applied severe shot peening (SSP) on AZ31 alloy with the aim to evaluate its potential in enhancing mechanical and electrochemical corrosion characteristics. As a low cost, versatile and flexible SPD treatment, SSP was developed by some of the authors, originally to enhance the fatigue properties of metallic materials commonly subjected to cyclic loading [41, 42]. The parameters that distinguish SSP from conventional shot peening are predominantly higher Almen intensity and surface coverage, which define the kinetic energy of the shot stream and the sample's exposure time, respectively. In SSP, the process parameters are set in a way to considerably increase the kinetic energy that is transmitted to the treated material in order to induce grain refinement. SSP has been applied to a wide class of materials including low alloy steels, stainless steels, cast iron and Al alloys [42-48]. This choice was motivated by reports on the favorable effects of conventional shot peening in improving limited fatigue properties of Mg alloys typically used in the automotive and aerospace industries [16, 29, 49, 50].

Interestingly, our previous studies also demonstrated the favorable effects of SSP on promoting cell-316L substrate interactions and reducing bacterial functions [51, 52]. Thus, herein, we considered the as-received (not peened) AZ31 samples as well as samples, which were shot peened with conventional parameters commonly used for this class of materials, as reference samples for controls. Structural and mechanical properties of the treated samples were evaluated through microscopical observations, surface morphology, roughness and wettability as well as microhardness and residual stress distribution measurements. X-ray photoelectron spectroscopy (XPS) analysis and electrochemical corrosion tests were executed to characterize the chemical composition and corrosion characteristics of the sample surfaces. Cytotoxicity tests with osteoblasts were also performed to ensure that the cytocompatibility of the material was not compromised by the SSP treatment. The present results showed for the first time the significant potential of SSP treatment to enhance the mechanical properties of Mg-based materials, like AZ31, with a limited effect on cytocompatiblity. The electrochemical characteristics were not improved after SSP due to the surface roughness and high density of structural defects induced on the surface during SSP. Samples that were subjected to slight surface grinding to reduce the surface roughness and separate its effect from that of the grain size, showed a high potential for decreasing this detrimental effect on material corrosion behavior. Further improvements are suggested to optimize the SSP process in order to reach its full potential to enhance functionality of Mg alloys in numerous orthopedic applications.

## 2. Experimental procedure
### 2.1. Material and surface treatment
Cold rolled and annealed AZ31 sheets of 6 mm thickness (Alfa Aesar GmbH, GE) were shot peened using the parameters described in Table 1. The parameters of the applied conventional shot peening (CSP) were chosen to represent the treatment commonly used for this class of material; severe shot peening (SSP) with much higher Almen intensity and surface coverage was considered to enhance the kinetic energy of the shot peening process in order to induce grain refinement on the top surface layer. The as-received not peened (NP) series was used as reference. Previous experimental data and the numerical model earlier developed by some of the authors [41] were utilized for the choice of SSP parameters. A repeened severe shot peened (RSSP) series, which were first SSP treated and subsequently subjected to a final soft repeening using glass peening media, were also considered with the intention of decreasing the surface roughness without affecting the surface microstructure after application of SSP treatment.

Table 1. Description of shot peening parameters

| Sample | Shot type | Almen intensity (mm) | Surface coverage% | Repeening Shot type | Repeening Surface coverage% |
|---|---|---|---|---|---|
| Not peened (NP) | - | - | - | - | - |
| Conventionally shot peened (CSP) | AZB100 | 0.15N | 100 | - | - |
| Severely shot peened (SSP) | AZB100 | 0.4N | 1000 | - | - |
| Repeened severely shot peened (RSSP) | AZB100 | 0.4N | 1000 | AGB6 | 100% |

AZB100 are ceramic beads composed of zirconium oxide, silica and alumina with a nominal size of 0.1-0.15 mm and a hardness of 640-780 HV. AGB6 are soda-lime glass beads with a nominal size and hardness ranging between 0.043-0.089 mm and 500-550 HV, respectively.

After the surface treatments, each plate was cut into discs 10 mm in diameter using a milling machine. Another complete batch of samples were prepared by slightly grinding all the shot peened sample treated surfaces. These samples were ground for a few seconds using 2500 grit size SiC paper just to slightly remove the surface morphology caused by shot peening. A series of NP samples were also ground to obtain the same surface roughness in all ground series. This procedure provided the ability to evaluate the effect of grain size without the interference of surface roughness. The ground set of samples are denoted in this paper by adding the letter "G" after their original series name (described in Table 1) and compared with the "as-treated" series. All the samples were rinsed ultrasonically in alcohol for 10 min before any experiment, unless specifically stated.

### 2.2. Microstructural analysis
In order to obtain a general understanding of the microstructural evolution in the surface layer and beneath it, samples of each series were cross sectioned, impregnated in cold mounting resin, ground with a series of SiC papers up to an average scratch size of 5 µm and then further polished using a polycrystalline diamond paste up to an average scratch size of 1 µm. The polished cross-sections were chemically etched using acetic-picral reagent (4.2 g of picric acid, 70 ml ethanol, 10 ml water and 10 ml acetic acid) and Nital 2% (2 ml nitric acid and 98 ml ethanol) sequentially for 5 seconds each. All etched samples were observed using a Leica DMLM light optical microscope (OM), in bright field mode. Another series of samples were prepared by positioning the samples with the treated surface facing up in the mounting resin; these series were also sequentially ground, polished and chemically etched following the abovementioned procedure, taking special precautions to remove minimum material from their surface. These latter samples were used to measure the grain size of the topmost surface layer of NP and CSP series, through image analysis of 100 individual grains from 3 samples per treatment using ImageJ software (National Institute of Health, US) [53].

Electron backscattered diffraction (EBSD) analysis of sample cross-sections was performed using a scanning electron microscope Tescan LYRA3XMU FEG/SEMxFIB equipped with X-Max80 EDS detector for X-ray microanalysis and EBSD detector Nordlys by Oxford Instruments with an Aztec control system (accelerating voltage of 20 HV). Specimen cross-sections were mounted in copper filled diallyl phthalate powder. The samples were ground and polished following the same procedure for samples of OM observation. The polished cross-sections were chemically etched using Nital 10 % (10 ml nitric acid and 90 ml ethanol) for 60-120 seconds. EBSD analysis was performed on an area of 100 × 200 µm² with the step size of 0.4 µm to cover the layer close to the treated surface as well as the less affected layer beneath it. The obtained data were processed using HKL Project Manager by Oxford Instruments determining the grain boundaries with minimum a mis-orientation angle larger than 10º.

Cross-section transmission electron microscopy (TEM) samples were prepared by mechanical polishing down to 30 µm, followed by ion milling in a Gatan PIPS machine at a 4 kV accelerating voltage and 7° incidence angle. Low-voltage (2 kV) milling was used as a final ion polishing step to reduce the amorphous surface layer that enveloped the specimen caused by ion milling. TEM observations were performed using a probe-corrected analytical high-resolution JEM ARM 200F electron microscope operating at 200 kV. The microscope was equipped with an IXRF EDS 2006 microanalysis system for energy dispersive X-ray spectroscopy (EDS).

### 2.3. Surface characterization

The surface morphology of the samples was analyzed and compared using a Zeiss Evo MA 15 scanning electron microscope (SEM). Surface micro-roughness parameters were quantified through surface analysis by an Alicona Infinite focus optical microscope with a profile length of 2 mm and a cut-off wavelength of 0.8 mm at 50× magnification. Five measurements were performed at randomly chosen areas on three samples per each series. Commonly used standard surface roughness parameters including arithmetic mean ($R_a$), root mean square deviation ($R_q$), and maximum height of the profile known also as peak to valley ($R_z$/PV) were measured. In order to study the symmetry and density of peaks of the profiles, two other parameters of skewness ($R_{sk}$) and kurtosis ($R_{ku}$) are also reported.

Surface wettability was determined using a Ramé-hart 100 optical contact angle goniometer. The water contact angle (WCA) between the surface line of the sample and the curved outline of 5 µl DI water droplets were measured on 2 random positions for each sample, using three samples per treatment. The measurements were performed at room temperature after a setting time of 5 s.

XPS experiments were carried out in an ultrahigh vacuum system with a base pressure of $1 \times 10^{-8}$ Pa, equipped with dual anode Mg/Al laboratory X-ray source (SPECS XR50) and electron energy analyzers (SPECS Phoibos 150). XPS spectra were recorded using an excitation energy of hν = 1253.6 eV (Mg Kα). In order to determine the thickness of an oxide overlayer, the samples were sputtered by Ar+ ions (1 keV), with the sputtering rate of 1.6 nm.hour$^{-1}$ (calibrated using attenuation of Mg 2p signal from the substrate, with the attenuation length as per TPP-2M formula [54]), in order to reveal the substrate contribution. Afterwards, the thickness was calculated from the angular variation of the ratio of the oxide and metal contributions to the Mg 2p signal [55].

### 2.4. Mechanical characterization

Microhardness measurements were performed on the lateral cross-sections of the samples after sequential grinding and polishing steps. HV0.25 indentations were produced by a Future-TECH FM-700 microhardness tester using a diamond Vickers indenter on three paths starting from the treated surface towards the core material.

X-ray diffraction analysis for residual stress measurements were performed in a AST X-Stress 3000 portable X-ray diffractometer using a CrKα radiation (Kα alpha = 2.2910 Å) source. The $\sin^2(\psi)$ method was used to analyze the data collected at a diffraction angle (2θ) of 152.4° corresponding to (104)-reflex scanned with a total of 11 Chi tilts in the range of − 45° to 45° along three rotations of 0°, 45° and 90° with a constant step size of 0.029°. Measurements were carried out on square samples of 40 x 40 $mm^2$. Instrumental broadening correction was applied using a reference strain free sample supplied by Proto XRD, CA. To obtain the profile of the residual stresses in depth, a thin layer of material (~ 0.03-0.04 mm) was removed from the surface of one sample per treatment at each step. A solution of acetic acid (94%) and perchloric acid (6%) was used in a Struers Lectropol-5 electro polishing instrument with a mask leaving a circular area of 3 mm in diameter of the sample surface exposed at each material removal step. The removal rate was measured between electro-polishing steps by a Mitutoyo micrometer precision IDCH0530/0560. Full width at half of the maximum intensity of the XRD peaks known as FWHM parameter was directly extracted from the XRD analysis.

**2.5. Corrosion properties**

The electrochemical corrosion tests were determined via a potentiostat VSP from BioLogic SAS, FR. The tests were carried out at room temperature using a classical three electrode system with platinum as a counter electrode, saturated calomel electrode (SCE) (+0.242 V vs. standard hydrogen electrode (SHE)) as a reference electrode and the sample as a working electrode [56]. Potentiodynamic polarization tests were carried out from -100 to + 200 mV vs SCE with respect to the open circuit potential (OCP) at a scan rate of 1 $mV.s^{-1}$ in a 0.9 % solution of NaCl, simulating the chloride concentration in the human body environment [57, 58]. The measurements were repeated three times per sample. The plotted potentiodynamic curves were analyzed using Tafel fitted by EC-Lab software.

**2.6. Assessment of Cytocompatibilty**

Samples were sterilized and cleaned by sequential ultrasonication in acetone and 200-proof ethanol, dried, and then submerged in 10 ml of complete DMEM for 3 days to prepare sample extracts. The samples were removed and the extracts were centrifuged briefly before being used in experiments. Primary human osteoblasts (PromoCell, Heidelberg, Germany) were cultured in Dulbecco's Modified Eagle's Medium (DMEM, Fisher) supplemented with 10% fetal bovine serum (FBS, ATCC) and 1% penicillin/streptomycin (P/S, Fisher) in a 37 °C, humidified, 5% $CO_2$/95% air environment. Cells at passage numbers of 3-6 were used in these experiments.

Osteoblasts were seeded into 96-well plates at a density of 5,000 cells/well and cultured for 1 day to allow the cells to acclimate. The media in each well was then replaced with sample extracts (and normal DMEM for the control group) and cultured under standard culture conditions for 1, 3, and 7 days. The cell media was changed every three days using the prepared extracts. After the prescribed time period, the media/extract was removed from each well and replaced with MTS reagent (Promega) mixed with complete DMEM in a 1:6 ratio according to the manufacturer's instructions. The MTS was developed for 3 h at 37 °C, allowing it to react with the metabolic products of viable osteoblasts, and then the absorbance of each well was measured at 490 nm. Experiments were performed in quadruplicate, and negative controls were employed to account for the absorbance of the sample extracts.

**2.7. Statistical analysis**

For all the experiments, the reported data refer to average values of all the measurements and their standard deviation, using three replicates per each group, unless specified. To evaluate the statistical significance of the obtained data, the analysis of variance (ANOVA) method followed by a Bonferroni's post-hoc test was used.

## 3. Results
### 3.1. Microstructural analysis

Top-view microstructural images of the sample surfaces are shown in Fig.1. The NP sample shows a regular pattern of fine equiaxed grains where the calculated average grain size was estimated as 15±2 µm. The CSP sample, Fig. 1 (b), represents visible deformation twins inside the equiaxed grains. The average grain size for the top surface of the CSP sample is measured to be about 11±2 µm. Importantly, these measurements were performed on the top surface layer of the samples after going through minimal grinding and polishing steps, and, thus, are not from the topmost layer, but the layer just under it. For SSP and RSSP samples, the extremely small grain size and high density of twin boundaries impeded individual grain size quantification after chemical etching, therefor, in that case, TEM analysis was performed to confirm grain refinement.

Lateral OM observations, Fig. 1 (c)-(f), clearly illustrate the microstructural evolution in the cross-sections of all shot peened samples. In the NP sample (Fig. 1 (c)), a single-phase structure alloy with even grain morphology and size was observed. For all the shot peened samples, on the other hand, the microstructure of the layer just beneath the surface was not distinguishable; this different etched top layer corresponds to a highly deformed zone, with grain sizes not observable by OM. The thickness of this dense top layer varied between the shot peened samples depending on the applied treatment, while the microstructure of the core material was comparable with that of the NP sample in all cases. The average thickness of this affected surface layer, measured at 30 different points on the images taken at 3 different surface areas per treatment, were determined to be about 35 µm for CSP and 65 µm for both the SSP and RSSP samples.

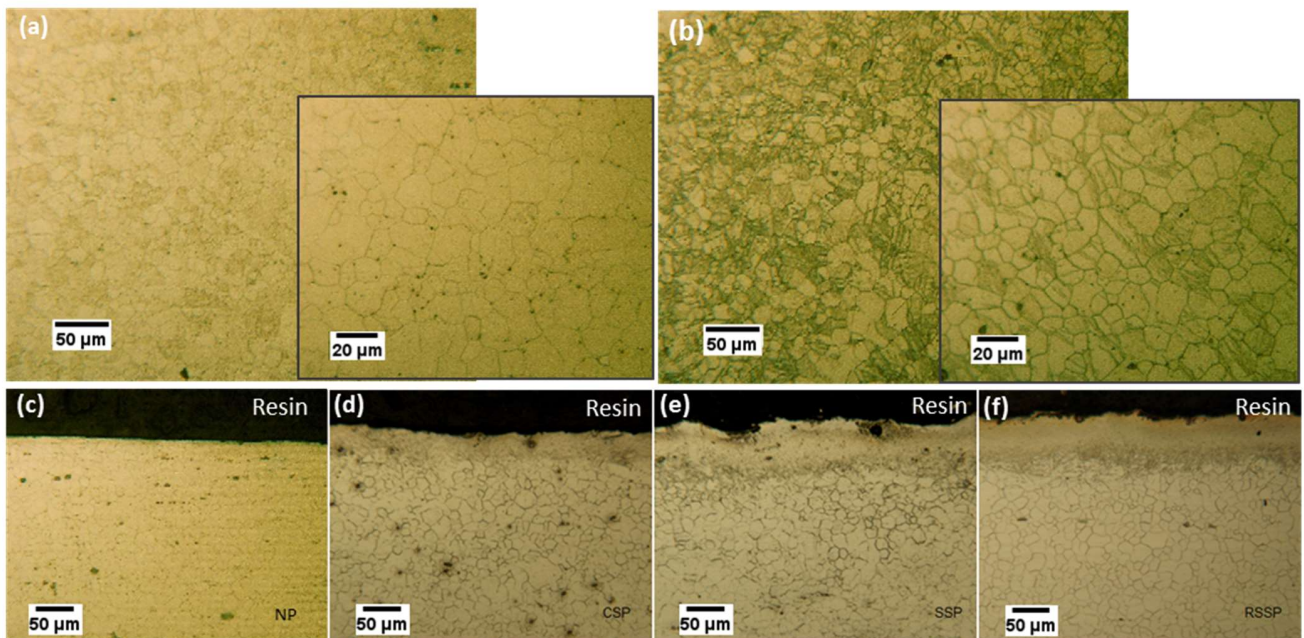

Fig.1. Representative optical microscopy micrographs of the top surface layer of (a) NP and (b) CSP samples at different magnifications. Cross sectional optical micrographs of (c) NP, (d) CSP, (e) SSP and (f) RSSP samples. NP: not peened, CSP: conventionally shot peened, SSP: severely shot peened and RSSP: repeened severely shot peened.

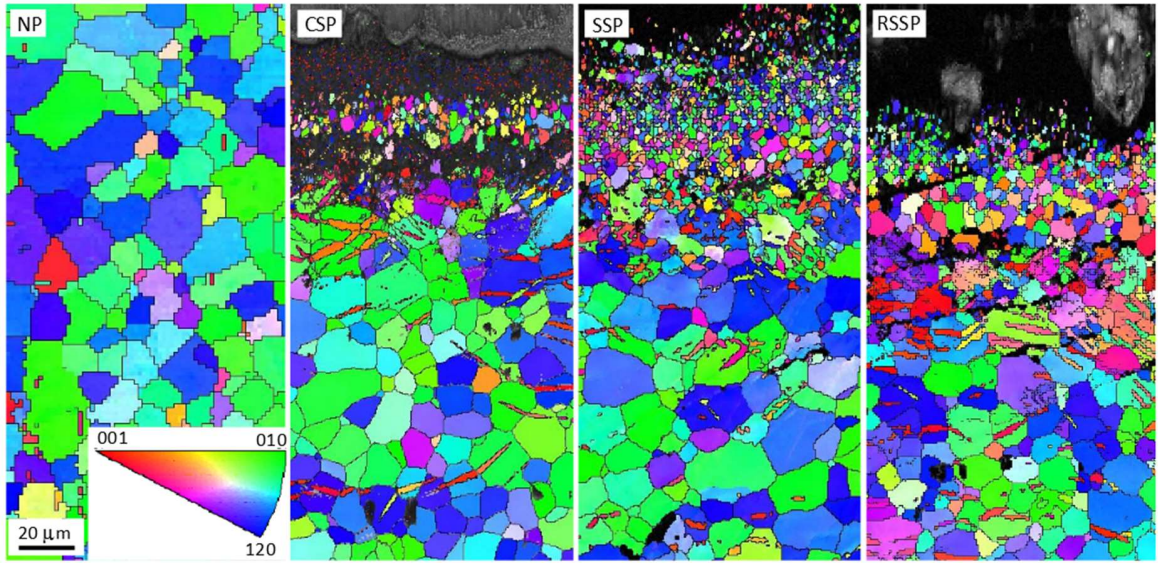

Fig. 2. EBSD maps of the lateral cross section of samples, (a) Internal reference from NP sample; near surface microstructure for (b) CSP (c) SSP and (d) RSSP samples. NP: not peened, CSP: conventionally shot peened, SSP: severely shot peened and RSSP: repeened severely shot peened.

EBSD maps, shown in Fig. 2, confirmed the OM observations, revealing notable grain refinement on the surface layers for all shot peened samples with a gradient grain structure, the evolution of which varied based on the applied treatment. The thickness of the notably grain refined layer was estimated to be around 45 μm for the CSP series and increased to around 80 μm for the SSP and RSSP series. Besides the grain size, the material texture was notably influenced by the shot peening treatment. A texture alteration from [010] to [120] was observed from EBSD maps of the internal reference (Fig. 2(a)) as well as for the material below the affected layer in the shot peened samples. All of the used treatment conditions resulted in a random orientation of newly formed fine grains in the shot peening affected area. Developed deformation twins can be also observed in the area below the highly refined layer, the density of which increased from the CSP sample to SSP and RSSP series. The black areas in the EBSD maps indicate a lower number of indexed points in the vicinity of the shot peened surfaces where the EBSD algorithm was unable to determine the orientation from the blurred or overlapping Kikuchi pattern. The geometry of the Kikuchi pattern is a projection of the crystal lattice and is unique for each particular structure and crystal lattice orientation.

In the bright field TEM observations of the NP sample (Fig. 3(a)), large grain parts are observed, represented by single spots on the corresponding selected area diffraction (SAD) pattern as shown in Fig. 3(b). These large grains contain nano-precipitates dispersed in the matrix as part of the original material structure; the nano-precipitates induced continuous rings in the corresponding SAD pattern. The EDS spectrum acquired from the NP sample in TEM mode is presented in Fig. 3(c). The peak identifications show the presence of Al, Mg and O attributing the nano-precipitates to an oxide compound of Mg and Al. The bright field image of the CSP sample (Fig. 3 (d)) indicates minor grain refinement, representing slightly refined grains with dimensions above 100 nm. The features observed on the surface originated from ion milling that created a trail of the original roughness through the material removal process. The spots in the SAD pattern (Fig. 3 (e)) correspond to large well-defined grains in the CSP sample. In Fig. 3 (f), on the topmost surface layer of the SSP samples, an amorphous layer containing dark areas representing random nano-precipitates can be observed. The corresponding SAD pattern with the hollow area in the center confirms the amorphous structure, and the continuous rings correspond to the nano-precipitates that are a part of the original structure, as also observed in the NP sample. This is confirmed by the interplanar distance between the high-resolution fringes of these precipitates, which correspond to Al, as shown in the SAD pattern. Just under the amorphous surface layer of the SSP sample, a mixed population of crystalline size was observed with larger grains containing small nanocrystals (<100 nm). SAD pattern analysis confirms that the nanostructures, causing the continuous rings, are Mg nanograins; this observation shows the effect of SSP treatment in inducing grain refinement. The individual spots on the rings correspond to the larger crystals. At a depth of 30 μm under the surface of the SSP sample, we can see a collection of individual grains with diameters higher than 100 nm. Getting further from the surface, at a depth of 150 μm, much larger grains were observed representing the typical structure of the alloy. Darker areas inside the large grains indicate the accumulation of strain and high deformation that can be attributed to the high plastic deformation exerted by the SSP treatment. A very similar structure to that of SSP was observed after RSSP treatment, as expected. Therefore, RSSP TEM observations are not included here for the sake of brevity.

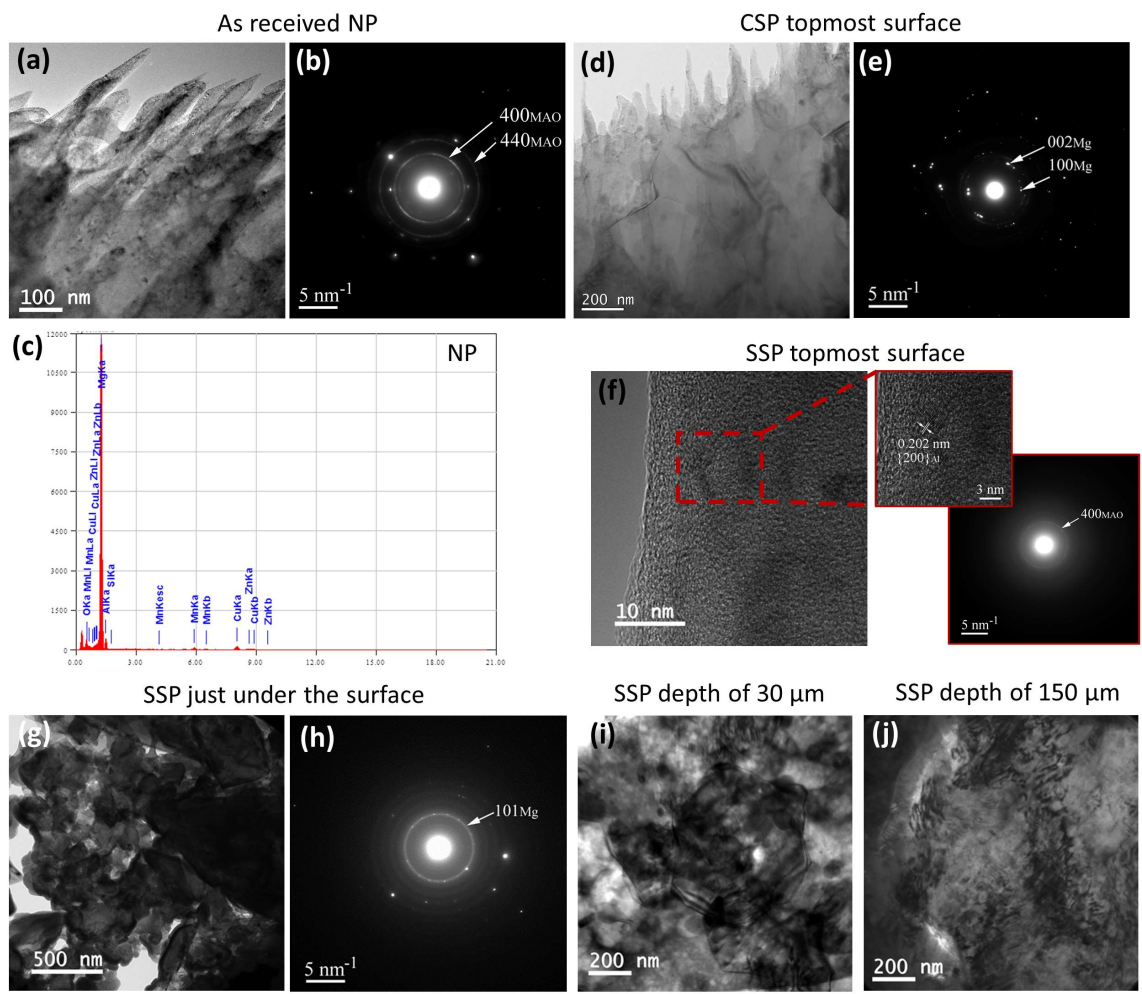

Fig. 3. TEM observations of the AZ31 samples: (a) Bright field image and (b) the corresponding SAED pattern of the NP sample; (c) EDS spectrum confirming the presence of Al, Mg and O in the nano-precipitates of the NP sample; (d) Bright field image and (e) the corresponding SAED pattern from the top surface of the CSP samples; (f) Bright field image from the topmost surface of the SSP samples (the inserts represent higher magnification bright field images of the nano-precipitates and its corresponding SAED pattern (MAO stands for $Mg_{0.36}Al_{2.44}O_4$, ASTM file No. 770729)); (g) Bright field image and (h) the corresponding SAED pattern of the SSP sample from just under the amorphous topmost layer; Bright field images from the depth of (i) 30 μm and (j) 150 μm of the SSP sample. NP: not peened, CSP: conventionally shot peened, and SSP: severely shot peened.

XPS was used to obtain an estimation of the elemental composition and, thus, assess the thickness of the oxide film at the surface of the samples after shot peening treatments. This thickness was estimated based on the chemical state information obtained on the surface layers after sputtering the sample surface with Ar$^+$ ions. In particular, for the NP sample, no metallic Mg2p signal was attained even after 6 hours of sputtering. Therefore, only the lower bound for the thickness of the oxide on NP was estimated. The thickness of the oxide layers were assessed to be > 186 Aº, 146 Aº, 95 Aº and 170 Aº for NP, CSP, SSP and RSSP samples, respectively.

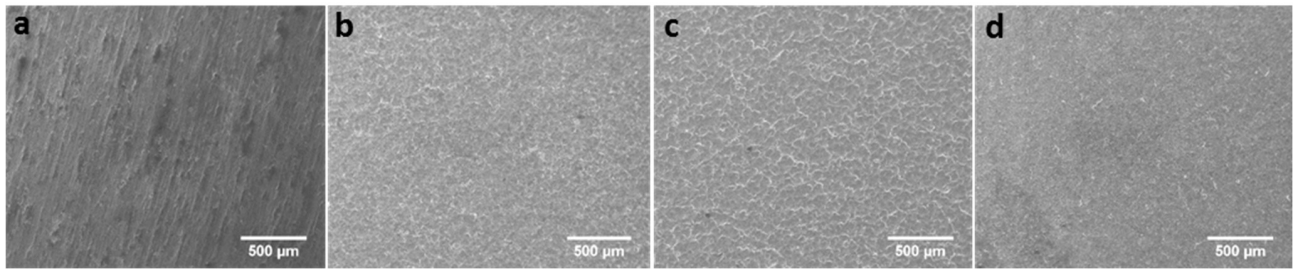

(e) 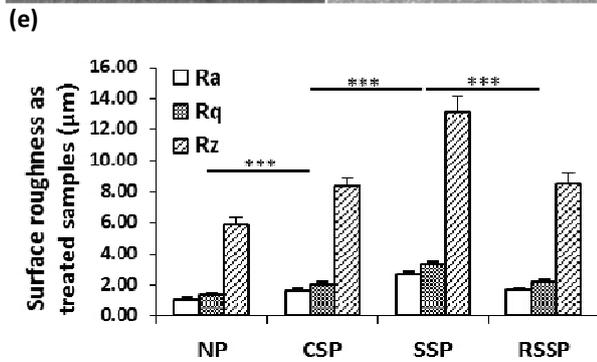 (f) 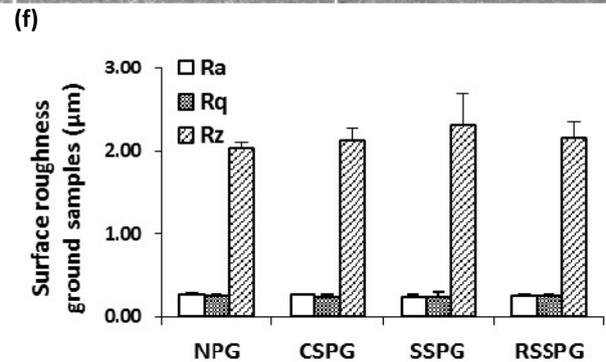

(g) 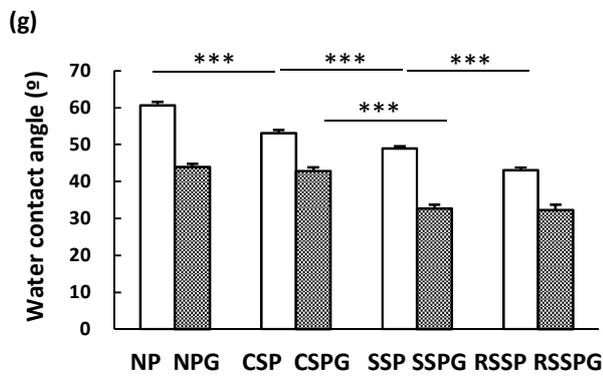 (h) 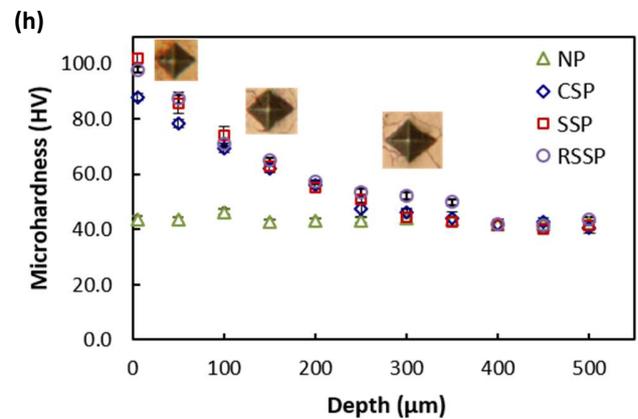

(i) (j)

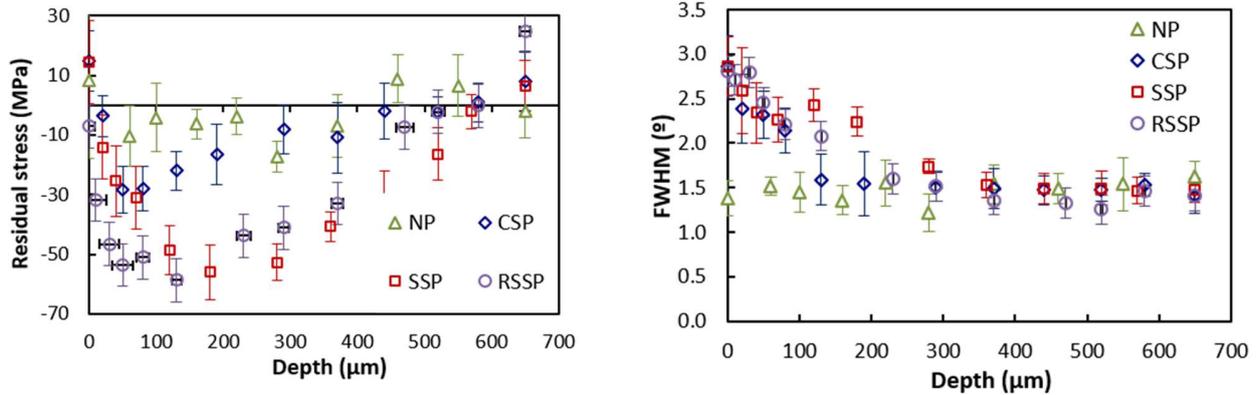

Fig.4. Top surface scanning electron micrographs of (a) NP, (b) CSP, (c) SSP and (d) RSSP samples; Surface roughness parameters measured on (e) as treated and (f) ground samples and (g) WCA data on as treated and ground samples; (h) Microhardness distribution on the sample cross section (the inserts are the optical micrographs of the indents at different depths on SSP sample); In-depth distribution of (i) residual stresses and (j) FWHM; NP: not peened, CSP: conventionally shot peened, SSP: severely shot peened and RSSP: repeened severely shot peened; Data = mean ± St. Dev.; N = 3, *** $p$ <0.001.

### 3.2. Surface characteristics

SEM micrographs of the NP sample shows a flat surface with some random scratches, while those of the shot peened series represent a rough surface with a chaotic texture typical for shot peened materials. CSP and RSSP samples displayed a relatively similar morphology. The SSP SEM observations, instead, revealed a higher density of visibly deeper surface indents caused by the multiple high energy impacts.

The data obtained in terms of surface roughness parameters for the as-treated samples, shown in Fig. 4 (e), vary significantly between different series, apart from CSP and RSSP that present no statistically significant difference in any of the measured parameters. The trend of $R_a$, $R_q$ and $R_z$ parameters are similar showing an increase in roughness from the CSP to SSP series.

Two additional surface roughness parameters of skewness and kurtosis were also quite similar on all of the as-treated series. Skewness is a measure of the symmetry of the profile describing the dominance of peaks or valleys. The skewness parameter was found to be very similar and close to zero for all samples, indicating that all the samples had a symmetrical surface topography with as many peaks as valleys. Similarly, the Kurtosis parameter was alike on all the samples and very close to 3. Kurtosis represents a key feature of the height distribution, namely the 'peakedness' of the profile. A surface with spiked height distribution has a kurtosis value greater than 3; while a surface that has a more low-tailed and broad spread height distribution would have a kurtosis value of less than 3.

The data obtained for the ground samples, Fig.4 (f), show no significant differences in terms of surface roughness parameters, demonstrating a similar topography obtained after grinding all the samples. This is essential to guarantee the suitability of the ground samples to highlight solely the effect of the micro/nanostructure on properties of the materials.

The results of the WCA measurements, presented in Fig.4 (g), indicated that a shot peening treatment in general promotes hydrophilicity of the AZ31 samples. For the as-treated series, a difference was seen in the contact angles by different shot peening treatments, showing the highest wettability for the RSSP series. In order to separate the contribution of surface roughness from that of grain refinement, the tests were conducted also on the ground series. The NPG and CSPG samples showed the same average contact angle and lower, but comparably similar data within the pair, was obtained also for the SSPG and RSSPG series. On the other hand, the results indicated an evident increase in wettability from the NP-CSP pair to the SSP-RSSP one, Fig.4 (g).

**3.3. Mechanical characteristics**

Microhardness profiles measured on sample cross-sections are presented in Fig.4 (h). The NP sample showed a constant microhardness typical of the base material that was around 43HV throughout the cross-section. The trends for all shot peened series depicted maximum microhardness close to the treated surface that gradually decreased to reach that of the base material at higher depths. No notable difference was observed on the maximum microhardness of CSP (88HV) compared to the other two series (100HV). However, amplifying the kinetic energy of the shot peening treatment increased the thickness of the affected layer; that is, the thickness of the layer displaying higher microhardness compared to the base material. This thickness was measured to be around 250 μm for the CSP samples, compared to around 350 μm and 400 μm SSP and RSSP, respectively.

XRD measurements provided the in-depth distribution of residual stresses and the FWHM parameter. The obtained in-depth residual stress profiles are shown in Fig.4 (i). Stresses for the NP series fluctuated around zero, as expected for the as-received material since it should not have noteworthy residual stresses. The two SSP and RSSP samples showed a similar distribution of compressive residual stresses. The residual stress distributions clearly varied between CSP and the two SSP series in terms of both maximum stress values (-30 MPa reached at a depth of 50 μm vs. -56 MPa reached at a depth of 180 μm) and the depth of annulment of compressive residual stresses (~300 μm vs. ~570 μm).

The FWHM parameter, extracted form XRD measurements, was influenced by lattice distortion, dislocation density, grain refinement, work hardening and instrumental broadening. The contribution of instrumental broadening was eliminated by correction using a strain free reference sample. Considering the NP sample as a control, the results presented in Fig. 4 (j) confirmed that all of the shot peening treatments introduced a substantial increase in FWHM close to the surface; this enhancement was observed to gradually decrease towards the core matrix. The important difference in the evolution of FWHM parameter was the depth at which the value reached that of the base material. This depth that can be considered as an estimation of the work hardened layer, which was measured to be around 130 μm for the CSP sample, 360 μm for SSP and 290 μm for RSSP.

**3.4. Corrosion tests**

Potentiodynamic polarization curves of AZ31 samples plotted on a semi-logarithmic scale are shown in Fig.5 for both the as-treated and ground series. The obtained kinetic and thermodynamic corrosion electrochemical characteristics complemented by the values of the Tafel constants ($\beta_a$ and $\beta_c$), expressing the slope of anodic and cathodic polarization curves, are listed in Table 2. Thermodynamic nobleness of the surface is expressed by more positive values of corrosion potential, $E_{corr}$. The results indicated that in all the tested series, shot peening slightly shifted the $E_{corr}$ to more positive values compared to the as-received NP series. The ground series exhibited a larger positive shift of $E_{corr}$ values. However, the measured $E_{corr}$ values were still very negative. The $i_{corr}$ values were reported in Table 2 for all the tested series, where much higher values were obtained for all shot peened samples when compared to the NP one.

Table 2. Electrochemical corrosion characteristics obtained by Tafel extrapolation analysis of the measured potentiodynamic curves

| Sample | $E_{corr}$ [mV] | $i_{corr}$ [µA.cm$^{-2}$] | $\beta_c$ [mV] | $\beta_a$ [mV] |
|---|---|---|---|---|
| NP | -1542 ± 25 | 21 ± 2 | 189 ± 10 | 66 ± 5 |
| CSP | -1503 ± 23 | 325 ± 5 | 144 ± 5 | 140 ± 6 |
| SSP | -1479 ± 26 | 326 ± 4 | 141 ± 6 | 144 ± 6 |
| RSSP | -1483 ± 27 | 321 ± 6 | 148 ± 6 | 145 ± 8 |
| NPG | -1508 ± 16 | 19 ± 2 | 180 ± 8 | 65 ± 4 |
| CSPG | -1492 ± 19 | 30 ± 3 | 225 ± 10 | 68 ± 2 |
| SSPG | -1448 ± 15 | 81 ± 5 | 290 ± 9 | 75 ± 4 |

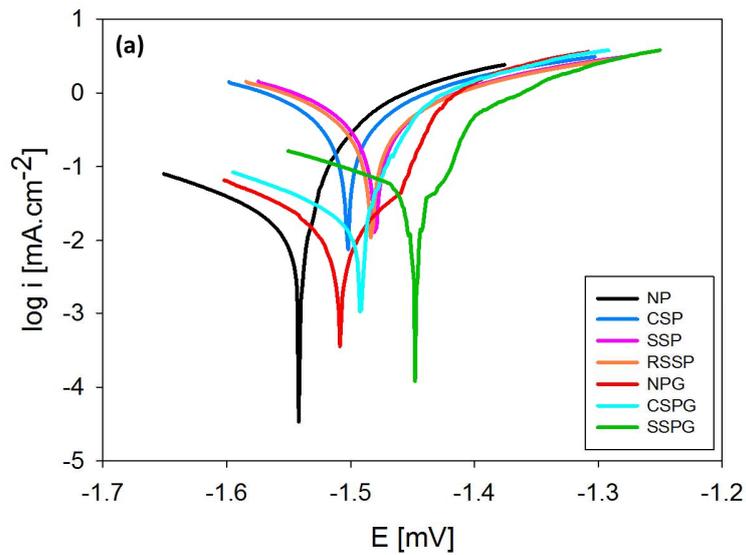

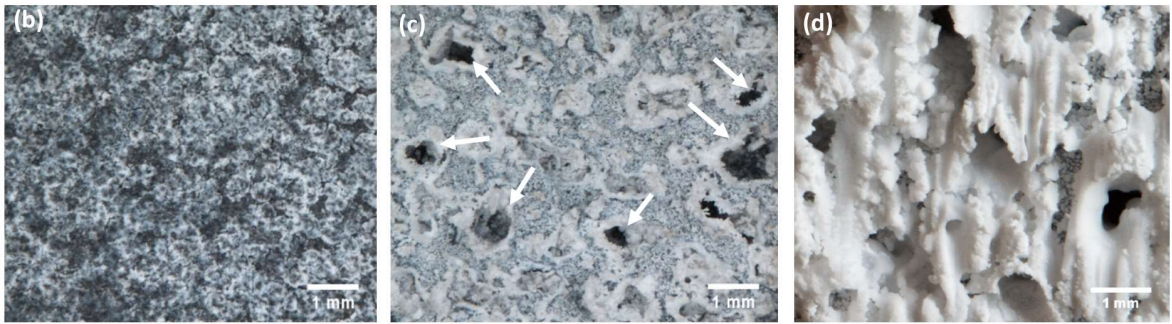

Figure 5. (a) Potentiodynamic curves of AZ31 samples in 0.9 % NaCl solution; Optical micrograph of the SSPG surface after (b) 15 min, (c) 17 h, and (d) 168 h immersion in 0.9 % NaCl solution at room temperature; the arrows indicate localized corrosion in correspondence with the residual craters caused by the high energy impacts. NP: not peened, CSP: conventionally shot peened, SSP: severely shot peened and RSSP: repeened severely shot peened.

The ground series, on the other hand, showed far lower values of $i_{corr}$, comparable to that of the NPG. Optical micrographs of the exposed surface of the SSPG surface shown in Fig. 5 (b)-(d) respectively after 15 min, 17 h and 168 h exposure times, marked the presence of dimples and craters caused by high energy impacts even after slight grinding. These craters seemed to have acted as local corrosion sites. One week of immersion in a corrosive environment led to extensive formation of magnesium corrosion products of almost 2 mm height from the original surface profile; it also resulted in intensive local dissolution inducing 1.5 mm deep craters in the most thermodynamically active regions.

### 3.5. Assessment of Cytocompatibility

Fig.6 shows the viability of osteoblasts cultured in sample extracts compared to control cells grown in normal DMEM. After 1 day of culture, cells grown in all media displayed roughly the same metabolic activity. While cells grown in the NP and CSP sample extracts maintained this trend, cells grown in the SSP and RSSP extracts exhibited decreased viabilities at day 3. However, after 7 days of culture, cells grown in the extract from all AZ31 samples displayed the same or slightly increased activities compared to the control. This indicates that, although diminished cellular activity was observed in the first three days of culture, there were no lasting cytotoxic effects of SSP samples on osteoblasts.

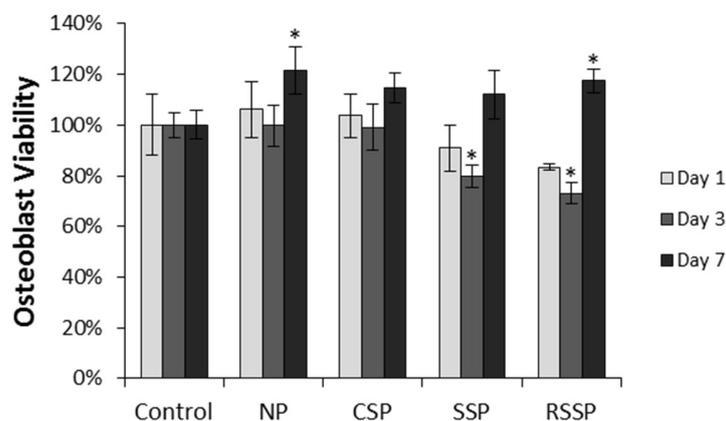

Figure 6. Viability of primary human osteoblasts cultured in sample extracts for 1, 3, and 7 days, as measured by an MTS assay. Data = mean ± St. Dev; N = 3, *p<0.01 compared to the control at the same time point. The overall activity of osteoblasts in each group statistically increased from day 1 to 3 to 7. NP: not peened, CSP: conventionally shot peened, SSP: severely shot peened and RSSP: repeened severely shot peened.

### 4. Discussion

Biodegradable Mg-based alloys have the potential to reduce the rate of cumbersome re-operation and retrieval surgeries by replacing non-degradable metallic materials commonly used in orthopaedic, trauma and maxillofacial surgery. However, there are still plenty of challenges to engineer mechanically and chemically functional biodegradable Mg-based implants. Metallurgical alloying and a wide range of surface coatings have been put forward to modulate the interaction of Mg materials with the biological environment; however, these methods have not been fully efficacious and there are still numerous challenges to be addressed. In the present study, AZ31 Mg alloy samples were treated by SSP, an impact based mechanical surface treatment that induces surface grain refinement, compressive residual stress fields, and an increasing surface roughness. Contrary to many other SPD based grain refining techniques, SSP is low cost and flexible in terms of size and geometry of the material to be treated. Our previous results also confirmed its efficacy in enhancing mechanical properties, promoting interaction of the treated surface with osteoblasts and reducing early bacterial adhesion on 316L samples [52]. We applied SSP to a series of AZ31 samples and their functionality was compared to two other shot peened series, treated with various process parameters (CSP and RSSP), in order to induce different combinations of grain size and surface roughness, as well as a reference as-received (NP) series. A grinding step was also performed on NP, CSP, SSP and RSSP series in order to isolate the effects of surface topography from grain refinement. The AZ31 alloy provides a good combination of chemical and mechanical characteristics and is primarily used for biomedical, aerospace and automotive applications. Its main alloying elements, as denoted by its designation, are aluminium and zinc with their respective weight concentration percentages of 3 % and 1 %.

The rationale behind this work was to apply this cost effective surface treatment to a biodegradable AZ31 alloy in order to simultaneously enhance its mechanical characteristics, modulate corrosion resistance, and verify the potential of SSP treatment in controlling the functionality of this alloy in the biological environment. Cytocompatibility tests were also performed using osteoblasts to evaluate the effect of various shot peening treatments on the cytocompatibility of the base material during degradation.

Microstructural OM top surface observations performed on NP and CSP samples revealed the general propensity of shot peening to induce surface grain refinement by indications of slight grain refinement even after CSP treatment compared to the NP material (11±2 µm vs. 15±2 µm). Considering that these measurements have been performed after grinding and polishing steps, which removed around 15-20 µm of the surface material, relatively more grain refinement should be expected on the treated surface of CSP samples, as validated by EBSD data. Furthermore, a high density of deformation twins were visible close to the surface after CSP treatment (Fig.1 (b)). A limited number of slip systems at room temperature is characteristic of hexagonal close packed (HCP) structures, such as Mg alloys, which predetermines the creation of deformation twins instead of dislocation slip under loading. Deformation twins, caused by high strains, introduce new boundaries within existing grains subdividing them into smaller areas and facilitating the grain refinement process [59]. It is to be noted that the high density of deformation twins on the top surface of SSP and RSSP samples impeded the possibility of measuring their surface grain size through OM image analysis; thus, EBSD and TEM were used to provide grain size information on these samples.

The results obtained from lateral cross section microscopical observations confirmed that increasing the kinetic energy of the process from CSP to SSP clearly increased the thickness of the affected surface layer, almost doubling it. This notable difference is a direct consequence of the higher Almen intensity used for the SSP treatments. RSSP samples have undergone a final re-peening process, just to reduce the surface roughness without changing the microstructure; hence, the RSSP samples are not expected to have a different surface structure when compared to SSP. The OM images confirmed the similarity of the microstructure and thickness of the affected layer for the two SSP and RSSP series.

EBSD mapping (Fig. 2) confirmed the OM observations, providing additional information about the shot peening affected areas and grain orientation within them. The results showed that shot peening treatment significantly reduced the grain size on a surface layer that was separated from the rest of the material matrix by sharp boundaries. TEM observations also confirmed smaller grain sizes in the top surface layer of CSP samples compared to NP, as well as nanograins in the surface layer of SSP samples. Both EBSD mapping and TEM analysis established gradual grain variation by depth in shot peened samples, starting from nanograins on the top surface and progressively increasing to submicron and then micron size traveling towards the core matrix. As EBSD maps exhibit, in the affected layer, the grains were significantly smaller compared to the core material. Fig. 2 illustrates that the extension of this grain-refined layer was found to increase with increasing kinetic energy of the treatment, from CSP to SSP and RSSP. These results are in agreement with our findings on the direct influence of SSP parameters on the thickness of the affected layer also on other materials [60]. The SSP treatment was also found to affect the texture that was a characteristic feature of the studied AZ31 magnesium alloy, as well as its grain orientation. The texture of the base material was altered in crystallographic orientations from [010] to [120] (Fig. 2). Random orientations of fine grains in the affected surface layer were observed on all shot peened series (Fig. 2). Even though there was a negligible effect of shot peening on grain size and grain orientation in the areas under the affected layer, the presence of deformation twins was quite notable in these regions. The density of deformation twins in the underlying structure increased by enhancing the kinetic energy of the applied treatment, similar to the trend observed for the extension of the affected layer (Fig. 2 (a)-(c)). Similar results for material response to plastic deformation was documented also by Amanov et al. [61] who treated AZ91D Mg alloy by a surface grain refinement technique called ultrasonic nanocrystalline surface modification. Evidence of accumulated strain and high deformation were observed at higher depths of around 150 μm in SSP samples (Fig. 3 (i) and (j)). Overall, the EBSD and TEM observations confirm that the extent of grain refinement within the surface layer and its depth varied linearly with the kinetic energy of the applied treatment.

Surface roughness is an important side effect of shot peening which can decisively influence the functionality of the treated material in many applications [62]. SEM observations and roughness measurements indicated that the surface morphology and surface roughness of the samples were notably changed after shot peening. SEM micrographs of SSP samples showed numerous overlapped indents and craters on its top surface; fewer defects were observed on the surface of the CSP and RSSP samples. However, it is interesting to note that the damage induced on the surface of all shot peened samples are fewer and less evident compared to the results reported for shot peening treatments on similar materials with similar Almen intensities but lower surface coverage [30], where a high density of pits was observed on treated surfaces even at lower Almen intensities. This difference can be attributed to the choice of the peening media.

During the shot peening process, despite the random sequence and impact location of the shots, a deformed surface with a relatively regular distribution of indents is obtained at high surface coverage. Homogenous final roughness is caused by the numerous overlapping plastic indentations; thus, it is common to use the most extensive standard surface roughness parameters to describe the surface topography of shot peened samples. The results of roughness measurements indicated a linear and direct relationship between the Almen intensity and the surface roughness parameters. The highest surface roughness was measured on the SSP sample. Slight repeening after SSP treatment (RSSP), resulted in a similar morphology and comparable roughness to the CSP series, despite the notable difference in Almen intensity. This similarity indicates the efficacy of the applied repeening treatment in smoothing the rough surface of the SSP series, rearranging the surface morphology after the SSP treatment. These results indicate that the high surface roughness induced by the high energy impacts of the SSP treatment, a potentially negative parameter for the fatigue strength, can be moderately decreased by slight repeening without affecting the positive effects of SSP, i.e. maintaining the compressive residual stresses and the nanograined surface layer. Both the skewness and peakedness parameters did not vary significantly with the shot peening treatments, signifying that the surface remains symmetric with a balanced distribution of peaks and valleys regardless of the process parameters.

Wettability tests determine how the differently treated surfaces would interact with the wetting media. This parameter is a key factor in protein absorption and subsequent interactions of the treated surface with cells [51]. The WCA measurements indicated a favorable effect of all shot peening treatments in enhancing the base material's wettability, with SSP samples showing the highest wettability. The WCA data were found to be inversely related to the surface roughness parameters. However, since the samples exhibited both different surface roughness and grain structure, this increased hydrophilicity could be due to both parameters. The influence of grain size was confirmed by the WCA measured on CSP and RSSP samples; as these two series have the same surface roughness, evidenced in Fig.4 (e), but their WCA are essentially different. This difference can be explained by the impact of different microstructures on WCA. To better understand the role of surface microstructure, WCAs were also measured on ground samples. Considering that surface roughness measurements confirmed similar topography on all ground series (Fig.4 (f)), the only difference between the WCA measured on these samples should be attributed to their surface grain structure. The results show similar WCA on NPG and CSPG samples; it can be due to the very limited thickness of the affected layer by CSP treatment, which was possibly removed after grinding and thus the exposed surface area in this case had the same microstructural characteristics of the NP sample. This is valid also for the SSPG and RSSPG samples, where the surface-restricted effect of repeening treatment was removed by the slight grinding; also, in this case, the similar roughness and similar grain structure resulted in similar wettability data on SSP and RSSP series. However, the statistically significant difference between the WCA data obtained for the NP-CSP pair and the SSP-RSSP pair can be attributed mainly to the different grain structures of their surface layers. This notable increase in surface wettability highlights the favorable effect of grain refinement on the interaction of the material with aqueous media.

XPS analysis provided quantitative data on the surface chemistry of each sample after shot peening treatment. These data were used to assess the thickness of the oxide film formed on the samples surfaces after shot peening. The XPS data indicated that the thickness of the oxide layer was the highest on the NP sample. The multiple impacts during the shot peening treatment seem to have dissected the original oxide overlay on the as-received NP material. The oxide layer was formed over the sample surface again after the shot peening treatment but its thickness was measured to be considerably thinner on the SSP sample. The RSSP samples, on the other hand, showed the highest oxide layer thickness among the shot peened series. The repeening treatment seems to have provided the possibility of increased oxidation on the RSSP sample surfaces. We postulated that the thickness of the oxide layer developed after shot peening could be influenced not only by surface roughness but also by the grain structure of the underlying layer, which directly affects the surface energy and reactivity. It was reported that ultra-fine grained AZ31 achieved via ECAP, that is, a SPD method used to obtain bulk grain refinement, offered more corrosion nucleation sites; however, the obtained ultra-fine structure led to easier and faster regeneration of the corrosion protective layer, which resulted in improved global corrosion properties [63].

Microhardness measurements confirmed the generally favorable effect of shot peening in work hardening the top surface layer (Fig. 4(h)). The extent of hardening was directly related to the kinetic energy of the applied shot peening treatment. The difference in microhardness of the SSP and RSSP series with respect to NP series is of particular interest, as it highlights a surface microhardness enhancement of more than 230% compared to the base material, as well as a thicker layer with higher microhardness compared to CSP series.

Compressive residual stresses are the key feature typically induced by shot peening. These stresses have a vital role in reducing crack propagation by keeping the crack faces pressed against each other (crack closure effect) and reducing their propagation rate under stress. XRD measurements showed the prominent effect of shot peening on the distribution of induced residual stresses. The measurements revealed higher compressive residual stresses and a greater depth of material affected by these stresses for the severely treated series (SSP and RSSP samples), highlighting the positive effect of increased kinetic energy in enhancing the compressive residual stresses (Fig. 4(i)). The difference in the compressive residual stress will certainly affect the material's fatigue strength. However, at the same time, increased surface roughness can dwindle the fatigue behavior by forming potential sites of stress concentration [43]. Moreover, the surface grain refinement can play a positive role in delaying crack initiation or change the crack initiation mechanism and, thus, promote fatigue strength enhancement [17, 42]. Therefore, the contribution of all these parameters is expected to enhance the fatigue strength of the SSP treated material, although further experiments are required to accurately estimate these effects.

FWHM evolution (Fig. 4(j)) also underlined the effects of the applied shot peening treatments in inducing lattice distortion, work hardening and grain refinement. The FWHM of shot peened samples exhibited the same trend as the microhardness. The depth of the layer showing higher FWHM values with respect to the base material was much greater for the SSP and RSSP series when compared to CSP. The Bragg reflection broadening after SSP treatments and the slight shift of peak centroid position were recognized as a result of micro strains and work hardening as well as grain refinement on the surface layers. Similar results for the SSP and RSSP samples were obtained, as expected, since the repeening treatment of the RSSP sample was not intense enough to induce any noticeable change with respect to SSP for both the compressive residual stress field and FWHM. The purpose of the repeening step was just to decrease the surface roughness without interfering with the other aspects of the material.

Potentiodynamic polarization curves indicated that shot peening treatments induced a slight shift in the negative corrosion potential, $E_{corr}$, of AZ31 samples to more positive values. This shift was more pronounced in the ground series. However, considering the standard deviation of the measured data and the clearly negative values of the determined corrosion potentials, the effect of this positive shift on general thermodynamic corrosion properties can be negligible. The minimal effect of shot peening on $E_{corr}$ was in agreement with [56], where a shift of 40 mV was determined even when the roughness of shot peened AZ31 changed from $R_a$ = 3.02 μm for the ground surface to $R_a$ = 12.33 μm for the shot peened surface. The corrosion current density $i_{corr}$, expressing the kinetics of ongoing corrosion processes in an electrolyte, can be considered a more relevant parameter for comparing the effect of various surface treatments on corrosion resistance of AZ31 samples [57]. This electrochemical characteristic has a direct relation with the corrosion rate and can be used to estimate the extent of corrosion in a specific electrolyte [64]. The corrosion current density results indicated that shot peening deteriorated the corrosion resistance of the base material, overall. For the treated series, $i_{corr}$ increased significantly on all shot peened surfaces when compared to the NP samples, regardless of the parameters used for shot peening. The higher electrochemical reactivity of the shot peened surfaces could be caused by the increased surface roughness and craters and defects induced in the top surface layers. Similar data in terms of reduced corrosion resistance of SMAT treated pure Mg and Mg–1Ca alloy were reported to be due to the high density of strain-induced crystalline defects in the top surface layer. An increased exposure time was reported to affect just the thickness of the grain refined layer and not the corrosion behavior of the SMATed samples [37].

The ground series, on the other hand, exhibited a significant decrease in corrosion current density when compared to the as-treated series, due to improved uniformity of the surface morphology. In this case, the determined $i_{corr}$ for the CSPG series reached values comparable to the NPG material. Slightly higher values were obtained for SSPG samples compared to NPG. Comparison between the $i_{corr}$ of the as-treated and the ground series highlights the remarkable role of surface roughness in reducing corrosion resistance from the kinetic point of view. Increased surface roughness is known to enlarge the exposed surface area, enhancing the active anodic surface and consequently reducing the corrosion resistance [56]. The subsequent slight grinding process reduced the surface roughness and significantly decreased the $i_{corr}$ of ground samples. The difference between the $i_{corr}$ values of CSPG and SSPG samples could be attributed to the reduced but still considerable presence of dislocation densities and microstructural defects that acted as corrosion attack active sites, resulting in higher reactivity and lower stability for SSPG sample surfaces. Considering the limited thickness of the grain refined layer, a very slight grinding was applied to avoid removing the whole grain refined layer. Indeed, after slight grinding, as observed in in Fig. 5 (b) and (c), there are still remaining signs of indentations on the surface of the SSPG sample. The optical micrographs of the SSPG surface over time indicates that these dimples provided localized sites of preferential corrosion initiation. We postulate that by optimizing the SSP treatment in a way to increase the thickness of the grain refined layer, it would be possible to entirely eliminate the surface indents by grinding or other surface machining technique, while still keeping a fine grained top layer. It is also interesting to note that there are many reports confirming that tensile residual stresses are detrimental to stress corrosion cracking resistance in Mg alloys, whereas compressive residual stresses are found to enhance stress corrosion cracking resistance due to the crack closure effect [65]. Therefore, putting aside the issue of enhanced surface roughness, SSP is expected to enhance the functionality of the material under stress corrosion conditions, considering that the presence of the compressive residual stresses induced by SSP can improve the resistance to crack initiation and propagation.

Apart from surface roughness and grain size, the crystallographic orientation of the grains are also supposed to play an important role in defining corrosion characteristics [66]. Experimental and theoretical studies have revealed that the corrosion resistance of the basal plane of AZ31 is higher than that of the other planes, thanks to its higher atomic coordination and thus lower surface energy; samples with decreased basal texture have shown lower corrosion resistance [67, 68]. Pu et al. [40] confirmed the significant influence of crystallographic orientations on corrosion resistance of grain refined AZ31, reporting that samples with little basal texture showed higher corrosion resistance compared to the series with strong basal texture, while grain size was held constant. Thus, further studies on crystallographic orientation are required to better explain the electrochemical corrosion characteristics of treated Mg alloys.

Table 3 summarizes the obtained results. Considering the combination of the investigated characteristics, RSSP treatment can be considered the optimal choice for the functionality of the treated material. RSSP series demonstrate similar grain size, thickness of grain refined layer, microhardness and residual stress distribution, corrosion characteristics and cytotoxicity to SSP series but represent better wettability and reduced surface irregularities.

Table 3. Summary of the data obtained from the characterization experiments

| | NP | CSP | SSP | RSSP |
|---|---|---|---|---|
| Microstructure | Original microstructure | Slight surface grain refinement for thickness of ~45 µm | Notable surface grain refinement for thickness of ~80 µm | Notable surface grain refinement for thickness of ~80 µm |
| Surface roughness (µm) | $R_a$=1.07±0.12 $R_q$=1.38±0.11 $R_z$=5.84±0.50 | $R_a$=1.63±0.12 $R_q$=2.03±0.15 $R_z$=8.35±0.54 | $R_a$=2.67±0.19 $R_q$=3.30±0.20 $R_z$=13.19±0.99 | $R_a$=1.68±0.11 $R_q$=2.14±0.16 $R_z$=8.49±0.68 |
| Surface wettability WCA (º) | As treated: 60.60±0.97 Ground: 43.95±0.86 | As treated: 53.10±0.89 Ground: 42.85±1.01 | As treated: 48.90±0.62 Ground: 32.70±1.10 | As treated: 43.05±0.74 Ground: 32.30±1.49 |
| Microhardness | Surface: 43.8±0.65 HV Thickness: - | Surface: 88.1±1.19 HV Thickness: 250 µm | Surface: 102±1.42 HV Thickness: 350 µm | Surface: 97.9±1.01 HV Thickness: 400 µm |
| Residual stresses | Max stress:-17±5 MPa Total depth: 370 µm | Max stress:-28.4±7 Total depth: 440 µm | Max stress:-56.3±6 Total depth: 570 µm | Max stress:-58.77±6 Total depth: 570 µm |
| $i_{corr}$ [µA.cm$^{-2}$] | As treated: 21±2 Ground: 19±2 | As treated: 325±5 Ground: 30±3 | As treated: 326±4 Ground: 81±5 | As treated: 321±6 Ground: - |
| Cytotoxicity | Negative | Negative | Negative | Negative |

On the subject of increasing the thickness of the affected layer, it is interesting to note that overpeening indications have been observed on Mg alloys at relatively low Almen intensities [69, 70]. This can be attributed to the limited deformability of the HCP crystal structure of Mg caused by a low number of active slip planes at room temperature. Grain refining after shot peening is primarily caused by twinning, the main plastic deformation mechanism of Mg alloys at low temperature [71]. It has been reported that for HCP crystal structure, non-basal slip systems would be activated at temperatures higher than 225 ºC [72, 73]. At these temperatures, the flow stress increases, resulting in higher total strain and thus enhanced deformability [74]. Warm conventional shot peening of wrought Mg alloy Mg–9Gd–2Y performed at 240 ºC was reported to induce higher work hardened depth, higher compressive residual stresses and less surface damage at the optimum Almen intensity of 0.15 mmN, when compared to shot peening at room temperature [75]. Performing SSP treatment at higher temperatures to activate more slip systems and amplify the effect of the treatment is the subject of a future study. This approach can result in a greater depth of grain refined layer and thus can provide the possibility of applying additional post treatments such as surface finishing to remove the majority of the surface micro defects without removing the whole grain refined surface layer.

## 5. Conclusions and outlook
AZ31 magnesium alloy samples were shot peened with different sets of parameters ranging from conventional treatment to severe shot peening to induce different combinations of grain size and surface roughness.

The results indicated that, despite the limited deformability of Mg alloy at room temperature, increasing the kinetic energy of the shot peening process induced notable grain refinement down to the nano regime on the top surface layer; in addition, the severe shot peening treatment increased the surface roughness, enhanced the microhardness and surface wettability, and induced compressive residual stresses in a deep surface layer. The trends of surface roughness, microhardness, residual stresses, FWHM, as well as extent of grain refinement and density of deformation twins, exhibited a direct relationship with the kinetic energy of the applied shot peening treatment. Repeening treatment after severe shot peening was found to be effective in reconfiguring the surface morphology and roughness with minimal changes to the other studied characteristics. Hence, the repeened severely shot peened series (RSSP) can be considered the optimal choice for the functionality of the AZ31 considering reduced risk of stress concentration and crack initiation, while showing greater wettability and better mechanical performance when compared to the other treatments. Slight surface grinding after shot peening treatments successfully separated the effect of surface roughness form that of grain refinement, highlighting the impact of grain size on promoting hydrophilicity of the samples' surfaces.

In all the cases, shot peening was found to decrease corrosion resistance in the AZ31 samples. Slight grinding after shot peening led to a considerable improvement in electrochemical corrosion characteristics. The increased rate of corrosion from SSP and RSSP samples, due to their higher surface roughness, led to a small decrease in osteoblast viability during the first three days of culture. However, the cells fully recovered by day seven, indicating that there was no lasting cytotoxicity induced by the degradation of the SSP treated samples.

In the future, we plan to increase the thickness of the affected grain refined layer through optimized warm severe shot peening treatment. A higher thickness of the grain refined layer can provide the possibility of eliminating the surface defects through grinding while keeping the nanograins at the exposed surface. The nanocrystallized surface with low surface roughness is expected to promote corrosion resistance, while simultaneously maintaining enhanced mechanical properties.

**Acknowledgements**

The authors declare no conflict of interests in this work. SB acknowledges funding from the European CERIC-ERIC Consortium for access to experimental facilities and financial support. Northeastern University funding is acknowledged. We also would like to thank Mr. Tomas Duchon of Charles University in Prague, Czech Republic for XPS measurements. Valuable discussions with Dr. Corneliu Ghica of National Institute of Materials Physics, Romania on TEM observations are gratefully appreciated.**References**

1. Perrone, G.S., et al., *The use of silk-based devices for fracture fixation.* Nature communications, 2014. **5**.
2. Togawa, D., et al., *Lumbar intervertebral body fusion cages: histological evaluation of clinically failed cages retrieved from humans.* J Bone Joint Surg Am, 2004. **86**(1): p. 70-79.
3. Van Dijk, M., et al., *The use of poly-L-lactic acid in lumbar interbody cages: design and biomechanical evaluation in vitro.* European Spine Journal, 2003. **12**(1): p. 34-40.
4. Singh Raman, R., S. Jafari, and S.E. Harandi, *Corrosion fatigue fracture of magnesium alloys in bioimplant applications: A review.* Engineering Fracture Mechanics, 2014.
5. Gu, X.-N. and Y.-F. Zheng, *A review on magnesium alloys as biodegradable materials.* Frontiers of Materials Science in China, 2010. **4**(2): p. 111-115.
6. Witte, F., *The history of biodegradable magnesium implants: A review.* Acta Biomaterialia, 2010. **6**(5): p. 1680-1692.
7. Tiwari, A. and A.N. Nordin, *Advanced Biomaterials and Biodevices.* 2014: John Wiley & Sons.20